\documentclass[natbib,preprint]{sigplanconf}
\usepackage{amsmath}
\usepackage{amssymb}
\usepackage{amsthm}
\usepackage{semantic}
\usepackage{listings}
\usepackage{stmaryrd}
\usepackage{todonotes}

\newtheorem{theorem}{Theorem}

\newtheorem{lemma}{Lemma}

\newtheorem{proposition}{Proposition}
\newtheorem{constraint}{Constraint}
\newtheorem{definition}{Definition}

\usepackage{color}
\definecolor{lightgray}{gray}{0.90}
\newcommand{\Gbox}[1]{\colorbox{lightgray}{\ensuremath{#1}}}

\lstdefinestyle{basic}{
showstringspaces=false,
language=C++,
columns=fixed,
basewidth=0.49em,
lineskip=0pt,
escapechar=@,xleftmargin=1pc,%
keywordstyle=\bfseries,
mathescape=true,%
basicstyle=\ttfamily,%
morekeywords={return,fix,var,proc,fun,func},%
}

\begin{document}

\newcommand{\funk}[0]{\ensuremath{\mathrm{F}^2\mathrm{C}}}

\newcommand{\kw}[1]{\texttt{#1}}
\newcommand{\code}[1]{\lstinline{#1}}
\newcommand{\dbr}[1]{\llbracket #1 \rrbracket_{\rho\sigma}}
\newcommand{\of}[0]{{:}}
\newcommand{\tsubst}[2]{\vec{#1}{:=}\vec{#2}}

\newcommand{\allt}[2]{\kw{<}#1\kw{>}\,#2}
\newcommand{\alle}[2]{\kw{<}#1\kw{>}\,#2}
\newcommand{\funty}[4]{\forall #1.\, #2 \overset{#3}{\to} #4}
\newcommand{\funt}[3]{\kw{func}\kw{(}#1,#3,\kw{[}#2\kw{]}\kw{)}}
\newcommand{\fune}[5]{\kw{fun}\kw{(}#1 \kw{:}#2\kw{)}\kw{:}#4 \kw{[}#3\kw{]} \{ #5 \}}
\newcommand{\funes}[5]{\kw{fun}\kw{(}#1 \kw{:}#2\kw{)}\kw{[}#3\kw{]}\{ #5 \}}
\newcommand{\funep}[5]{\kw{fun}\kw{(}#1 \kw{:}#2\kw{)}#3\kw{:}#4 \{ #5 \}}
\newcommand{\letrec}[9]{\kw{fun}\,[#1]\; #2\kw{(}#3 \kw{:}#4,#6\kw{)}\kw{:}#5\,#9 \{ #7 \}\; #8}
\newcommand{\letrecbb}[9]{\mathsf{fun}\, #2(#3,#6) \{ #7 \}\; #8}
\newcommand{\LET}[3]{\kw{var}\; #1 \kw{=} #2\kw{;}\; #3}
\newcommand{\LETE}[3]{\kw{let}\,#1 \kw{=} #2 \,\kw{in}\, #3}
\newcommand{\letcall}[5]{\kw{let}\,#1 \kw{=} #2 \langle #3 \rangle\kw{(}#4\kw{)}\,\kw{in}\; #5}
\newcommand{\letcal}[4]{\kw{var}\; #1 \kw{=} #2 \kw{(}#3\kw{)}\kw{;}\; #4}
\newcommand{\letcallbb}[5]{\kw{let}\,#1 \kw{=} #2\kw{(}#4\kw{)}\,\kw{in}\; #5}
\newcommand{\ret}[1]{\kw{return}\,#1\kw{;}}
\newcommand{\tailcall}[3]{\kw{return}\,#1 \langle #2 \rangle \kw{(}#3\kw{);}}
\newcommand{\tailcal}[2]{\kw{return}\,#1 \kw{(}#2\kw{);}}
\newcommand{\tailcallbb}[3]{\kw{return}\,#1\kw{(}#3\kw{);}}
\newcommand{\locals}[1]{\mathit{locals}\llbracket #1 \rrbracket}
\newcommand{\clos}[8]{\langle[#1] #2(#3,#4)\,#5\,#6, #7\rangle}
\newcommand{\closbb}[7]{\langle #2(#3,#4)\,#6, #7 \rangle}
\newcommand{\INT}[0]{\kw{int}}
\newcommand{\LIST}[1]{#1\, \kw{list}}
\newcommand{\BOOL}[0]{\kw{bool}}
\newcommand{\cond}[3]{\kw{if}\,#1\,\kw{then}\,#2\,\kw{else}\,#3}
\newcommand{\fst}[1]{\kw{fst}(#1)}
\newcommand{\snd}[1]{\kw{snd}(#1)}
\newcommand{\pair}[2]{(#1, #2)}
\newcommand{\subst}[2]{\vec{#1}{:=}\vec{#2}}
\newcommand{\appsubst}[3]{[\vec{#1}{:=}\vec{#2}]#3}
\newcommand{\subs}[3]{[#1{:=}#2]#3}
\newcommand{\efapp}[2]{#1\kw{<} #2 \kw{>}}
\newcommand{\efappr}[2]{#1[#2]}
\newcommand{\toreg}[2]{\llbracket#1\rrbracket_{#2}}
\newcommand{\relate}[3]{#1 \sim_{#2} #3}
\newcommand{\erase}[1]{\lfloor #1 \rfloor}

\title{Effects for Funargs}

\authorinfo{Jeremy G. Siek, Michael M. Vitousek, and Jonathan D. Turner}
           {University of Colorado at Boulder}
           {jeremy.siek@colorado.edu}



\maketitle

\begin{abstract}

Stack allocation and first-class functions don't naturally mix
together. In this paper we show that a type and effect system can be
the detergent that helps these features form a nice emulsion. Our
interest in this problem comes from our work on the Chapel language,
but this problem is also relevant to lambda expressions in C++ and
blocks in Objective C. The difficulty in mixing first-class functions
and stack allocation is a tension between safety, efficiency, and
simplicity. To preserve safety, one must worry about functions
outliving the variables they reference: the classic upward funarg
problem. There are systems which regain safety but lose
programmer-predictable efficiency, and ones that provide both safety
and efficiency, but give up simplicity by exposing regions to the
programmer. In this paper we present a simple design that combines a
type and effect system, for safety, with function-local storage, for
control over efficiency.

\end{abstract}

\section{Introduction}

This paper describes a design for integrating first-class functions
into languages with stack allocation in a way that does not compromise
type safety or performance and that strives for simplicity.  This
design is intended for use in the Chapel programming
language~\citep{Chamberlain:2005fd}, but could also provide a safer
alternative to the new lambda expressions of C++~\citep{ISO:2011uq}
and a more efficient alternative to the blocks of Objective
C~\citep{Apple:2011fk}.  The design is meant for performance-oriented
languages in which the run-time overhead for each language construct
should be relatively small and predictable.

The straightforward integration of first-class functions into a
language with stack allocation poses type safety problems because of
the classic \emph{upward funarg problem}~\citep{Moses:1970fk},
illustrated in Figure~\ref{fig:upward} in the Chapel language. The
\code{compose} function returns an anonymous function which refers to
parameters \code{f} and \code{g} of the surrounding \code{compose}
function.  Both \code{f} and \code{g} are functions of type
\code{func(int,int)}; the first \code{int} is the input type and
the second \code{int} is the return type. The example then defines the
function \code{inc} and invokes \code{compose} to obtain \code{inc2},
a function that increments its argument twice.  Because Chapel
allocates parameters and local variables on the call stack, the
variables \code{f} and \code{g} are no longer live when \code{inc2} is
called; the call to \code{compose} has completed. During the call to
\code{inc2}, the locations previously allocated to \code{f} and
\code{g} may contain values of types that are different from
\texttt{func(int,int)}, so we have a counterexample to type safety.

\begin{figure}[tbp]
  \centering
\begin{lstlisting}
proc compose(f: func(int,int),
             g: func(int,int)): func(int,int) {
   return fun(x:int){ var y=f(x); return g(y); };
}
proc inc(x: int): int { return x + 1; }
var inc2 = compose(inc, inc);
inc2(0);
\end{lstlisting}
  \caption{Example of an upward funarg in Chapel.}
  \label{fig:upward}
\end{figure}

To avoid the upward funarg problem, most languages with first-class
functions do not allocate parameters or local variables on the stack;
they allocate them on the heap and use garbage collection to reclaim
the memory.  However, designing garbage collection algorithms that
provide predictable performance is an on-going research challenge
whereas stack allocation is reliably fast~\citep{Miller:1994fk}.
Advanced compilers for functional languages employ static analyses to
determine which variables may be safely allocated on the
stack~\citep{Guy-L.-Steele:1978yq,Goldberg:1990pi,Tofte:1994uq,Serrano:1996uq},
which improves efficiency for many programs, but still does not
deliver \emph{programmer predictable} efficiency~\citep{Tofte:2004fk}.

For the Chapel programming language, we seek a language design in
which upward funargs are caught statically, thereby achieving type
safety, but that enables higher-order programming and gives the
programmer control over run-time costs.

While such a design is not present in the literature, the key
ingredients are.  \citet{Tofte:1994uq,Tofte:1997fk} designed a typed
intermediate language, which we refer to as the Region Calculus, with
an explicit \emph{region} abstraction: values are allocated into
regions and regions are allocated/deallocated in a LIFO fashion.  The
Region Calculus uses a type and effect system \emph{\`a la}
\citet{Talpin:1992vn} to guarantee the absence of memory errors, such
as the dangling references that occur within upward funargs.
Later work relaxed the LIFO restriction on allocation and
deallocation~\cite{Aiken:1995fk,Crary:1999fk,Henglein:2001zr}.
\citet{Tofte:2004fk} suggest that a programming language (as opposed
to an intermediate language) with explicit regions would provide both
predictable efficiency and type safety.  The work by
\citet{Grossman:2002cr} on Cyclone provides evidence that this is the
case.

While the type and effect systems of \citet{Tofte:2004fk} and
\citet{Grossman:2002cr} support many higher-order programming idioms,
they still disallow many useful cases that involve curried functions,
such as the above \code{compose} function. The lambda expressions of
C++~\citep{ISO:2011uq} and the blocks of Objective
C~\citep{Apple:2011fk} offer a simple solution: enable the copying of
values into extra storage associated with a function.  For example, in
the \code{compose} function, the programmer could elect to copy the
values of \code{f} and \code{g} into storage associated with
the anonymous function, thereby keeping them alive for the lifetime of the
function. One might be tempted to make this copying
behavior the only semantics, but in many situations the copy is too
expensive~\citep{Jarvi:2007fk}. (Suppose the copied object is an
array.) In the C++ design, the programmer may choose to either make
the copy, incurring run-time cost in exchange for safety, or capture
the reference, incurring no extra run-time cost but exposing
themselves to the potential for dangling references.

We take away the following points from this prior research: 
\begin{enumerate}
\item The Region Calculus demonstrates that a type and effect system
  can support many higher-order programming idioms while disallowing
  upward funargs.
\item Cyclone shows that only a small amount of annotations are
  needed to support a type and effect system.
\item The C++ and Objective C approach of providing function-local
  storage enables the full spectrum of higher-order programming while
  keeping the programmer in control of run-time costs.
\end{enumerate}

In this paper we present a language design that uses a type and effect
system to detect and disallow upward funargs with dangling references
and that also offers the ability to copy values into function-local
storage.  However, unlike the Region Calculus and Cyclone, we do not
expose the region abstraction to the programmer; doing so would
unnecessarily complicate the language from the programmer's viewpoint.
In our system, the effect of an expression is a \emph{set of
  variables} instead of a set of region names. The variables in the
effect are those that may be read by the expression.

For readers unfamiliar with regions, we present a static and dynamic
semantics directly for our system and a syntactic proof of type safety
(Section~\ref{sec:direct}). The goal of this section is to provide
both an aid to implementation and a direct understanding of why our
type and effect system ensures type safety, and therefore also memory
safety~\cite{Pierce:2002hj}. As an added benefit, the direct semantics
supports efficient tail calls through a simple restriction in the type
system that enables early deallocation.
For readers familiar with regions, we present a type-preserving
translation from well-typed terms of our system into the Region
Calculus (Section~\ref{sec:translate-to-regions}).

This paper makes the following contributions:
\begin{enumerate}
\item We make an explicit connection between type and effect systems
  and the upward funarg problem so that implementers adding
  first-class functions to stack-based languages, such as C++,
  Objective-C, and Chapel, can leverage the wealth of prior work on
  effects.

\item We present a design that balances simplicity (for the programmer
  and the implementer) with safety and performance.  We formalize the
  design in the definition of a calculus named Featherweight
  Functional Chapel (\funk{}). An interpreter and type checker for
  \funk{} are at the URL in the supplemental material.

\item We give a direct proof of type safety for \funk{}, show that
  \funk{} is parametric with respect to effects by constructing an
  erasure-based semantics, and we relate \funk{} to the Region
  Calculus through a type-preserving translation.
\end{enumerate}
Section~\ref{sec:related-work} places our contributions in the context
of the prior research in this area and Section~\ref{sec:conclusion}
concludes the paper.


\section{Overview of the Design}
\label{sec:overview}

\begin{figure}[tbp]
  \centering
\[
\begin{array}{lrcl}
  \text{variables} & x,y \\ 
  \text{integers} & n & \in & \mathbb{Z}\\
  \text{operators} & \mathit{op} & ::= & + \mid - \mid \ldots \\
  \text{effects} & \varphi & ::= & x, \ldots, x  \\
  \text{types} & T & ::= & \INT{} 
                           \mid \funt{T}{\varphi}{T} \mid \allt{x}{T} \\
  \text{expressions} & e & ::= & x \mid n \mid  \mathit{op}\kw{(}e,\ldots,e\kw{)} \mid f \\ 
    & & & \mid \LETE{x}{e}{e} \mid \kw{fix}\,x\of T.\,f \mid \efapp{e}{x}\\
  \text{abstractions} & f & ::= & \funes{x}{T}{\varphi}{T}{s} \mid \alle{x}{f} \\
  \text{statements} & s & ::= & \LET{x}{e}{s} \mid \ret{e} \\
                    &   &     & \mid \letcal{x}{e}{e}{s} \mid \ret{e(e)} 
\end{array}
\]
  \caption{The syntax of \funk.}
  \label{fig:surface-syntax}
\end{figure}

We start with an overview of our design.
To communicate the design as clearly and succinctly as possible, we
present a small higher-order, stack-based calculus, named \funk.  The
calculus has explicit type and effect annotations. We point readers
interested in removing the annotations to the techniques developed for
Cyclone~\citep{Grossman:2002cr}.  We first describe the syntax for
\funk{}, defined in Figure~\ref{fig:surface-syntax}, and then present
the important design decisions through a series of examples.

The syntax of \funk{} is separated into expressions and statements to
streamline the dynamic semantics.\footnote{For the monadically
  inclined reader, the split almost puts \funk{} into monadic style
  with the statements in a monad for the call stack.  For
  more on monadic regions, we refer the reader to \citet{Fluet:2006hb}.
  }  This syntax
can be viewed as a variant of A-normal form and there are standard
techniques for translating programs into this form~\cite{Flanagan:1993cg}.

Expressions do not contain function calls or anything else that can
change the stack. The expressions do include many standard things such
as variables, integer literals, and primitive operator application.
We sometimes use infix notation for operator application. For example,
\code{x + z} should be read as \code{+(x,z)}. More importantly, \funk{} has a
\textit{function creation} expression:
\[
  \funes{x}{T}{\varphi}{T_2}{s}
\]
The $x$ is the function's parameter, of type $T$.  In examples, we
sometimes take the liberty of using functions with more than one
parameter. 
The effect annotation $\varphi$ declares the set of variables from
surrounding scopes that may be referred to in the body $s$ of the
function.  The type of the above function is
$\funt{T}{\varphi}{T'}$ where $T'$ is the return type of the
statement $s$. The effect $\varphi$ in the function's type is used to
make sure that the function is only called when all the variables in
$\varphi$ are live.

There is also a \code{let} expression that is completely standard but
plays an important role in modeling function-local storage, to be
discussed shortly.
The \code{fix} expression is also standard, and enables writing
recursive functions. We include \code{fix} in \funk{} primarily to
make sure our design does not accidently rely on the assumption that
all expressions terminate.
We postpone discussing the remaining two forms of expressions.

The statements of \funk{} are also defined in
Figure~\ref{fig:surface-syntax}. We syntactically enforce that every
control-flow path ends with a return statement to avoid obscuring the
type system with the standard machinery for preventing functions from
falling off the end~\cite{Grossman:2002cr}.
The \emph{variable initialization} statement
\[
\LET{x}{e}{s}
\]
allocates a new location on the stack with the result of $e$, binding
the address to the variable $x$, then executes statement
$s$ with $x$ in scope. 
The \code{return} statement is standard.

Statements include two forms of function call. The \emph{function
  call}
\[
\letcal{x}{e_1}{e_2}{s}
\]
invokes the function resulting from $e_1$ with the argument resulting
from $e_2$. The return value is placed on the stack and its
address is bound to $x$, which can then be used in the subsequently
executed statement $s$. The effect in the function type of $e_1$ must only
contain variables that are live in the current scope.
The \emph{tail call} 
\[
\tailcal{e_1}{e_2}
\]
is a function call in which nothing else is left to be done in the
current function after the call $e_1\kw{(}e_2\kw{)}$ returns.

\paragraph{Allow Downward Funargs}

Several variations on the example in Figure~\ref{fig:twice} serve to
demonstrate the interplay between first-class functions, stack
allocation, and effect annotations.  The example defines a
\code{twice} function, with a function parameter \code{f} that it
calls twice, first on the parameter \code{y} and then on the result of
the first call. The example also defines the \code{addx} function,
which adds its parameter \code{z} and the global variable \code{x}.
The example then invokes the \code{twice} function with \code{addx} as
a parameter; so the \code{addx} argument is an example of a downward
funarg.

\begin{figure}[tbp]
  \centering
\begin{lstlisting}[numbers=left]
var x = 1;
var addx = fun(z: int)[x]{ return x + z; };
var twice = fun(f: func(int,int,[x]), y: int)[x] {
               var t = f(y); return f(t);
            };
var b = twice(addx, 3);
return b;
\end{lstlisting}  
  \caption{Example with first-class functions and stack allocation.}
  \label{fig:twice}
\end{figure}

Downward funargs are benign with respect to memory safety and are
allowed by our type system.  In this case, the \code{addx} function
reads from \code{x}, so \code{x} is recorded in the type of
\code{addx}. The call to \code{twice} with \code{addx} is allowed by
the type system because the type of \code{addx}, including its effect,
matches the type of parameter \code{f}. The call to \code{f} inside
\code{twice} is allowed because \code{twice} has also declared
\code{x} as its effect.  Then looking back to the call to
\code{twice}, it is allowed because the lifetime of variable \code{x}
encompasses the call to \code{twice}.

However, the above \code{twice} function is surprisingly specific. It
may only be called with a function that reads from \code{x}. We
discuss shortly how to make \code{twice} more general.

\paragraph{Disallow Upward Funargs}

Next consider the \code{twice} function written in curried form,
that is, taking one argument at a time as shown below.
The function created on line 4 is an example of an upward funarg that
is disallowed by our type system.  It reads from three variables, one
that is local to the function (\code{y}) and two from surrounding
scopes (directly from \code{f} and indirectly from \code{x}).  The
return of this function is disallowed because the variable \code{f} is
going out of scope, so this upward funarg contains a dangling
reference.
\begin{lstlisting}[numbers=left,firstnumber=3]
var twice = fun(f: func(int,int,[x])) {
     return fun(y:int)[f,x] {
               var t = f(y); return f(t);
            };             /* Error */
   };
\end{lstlisting}

\paragraph{Function-local Storage}

In \funk{}, the programmer can solve the upward funarg problem by
copying \code{f} into function-local storage. This can be
accomplished with a \code{let} expression, which copies (via
substitution) the result of its right-hand-side into its body.  With
this change, the function no longer reads from \code{f} when it is
called (the read occurs in the \code{let}, during the execution of
\code{twice}), so the type system allows the return of this upward
funarg from \code{twice}.
\begin{lstlisting}[numbers=left,firstnumber=3]
var twice = fun(f: func(int,int,[x])) {
    return (let g=f
            in fun(y:int)[x] {
                  var t = g(y); return g(t);
               });
  };
\end{lstlisting}
To see how this works, consider the following step of execution, where
$\mathit{addx}\equiv \funes{z}{\INT}{x}{}{\ret{x + z}}$.
\begin{align*}
    & \; \LETE{g}{\mathit{addx}}{\funes{y}{\INT}{x}{}{ \LET{t}{g(y)}{\tailcal{g}{t}}}} \\
\longrightarrow & \; \funes{y}{\INT}{x}{}{ \LET{t}{\mathit{addx}(y)}{\tailcal{\mathit{addx}}{t}}}
\end{align*}
The \code{let} expression has caused the \textit{addx} function to be
copied into the body of the function.

Of course, a production quality compiler would not use substitution
(which implies run-time code generation) but instead would use a
closure representation of functions. A closure consists of a function
pointer paired with an array of the values from the \code{let}-bound and
\code{fix}-bound variables that occur free in the function body.  For readers
familiar with C++, a closure is just a ``function object'' or
``functor'' in which data members are used to store copies of the
variables.

To reduce the notational overhead of adding \code{let} expressions,
one can add syntactic sugar to function expressions for declaring that
a variable be copied, as in C++~\citep{ISO:2011uq}. The following
shows \code{f} listed after the normal parameters of the function on
line 4 to indicate that \code{f} is to be copied into function-local
storage.
\begin{lstlisting}[numbers=left,firstnumber=3]
var twice = fun(f: func(int,int,[x])) {
     return fun(y:int; f)[x] {
               var t = f(y); return f(t);
            };
  };
\end{lstlisting}

\paragraph{Effect Polymorphism}

As described so far, the type and effect system is too restrictive
because function types are annotated with specific variables.  For
example, the below \code{twice} function has a function parameter
\code{f} that is annotated with the effect \code{[x]}. Thus, calling
\code{twice} with \code{addx} is fine but calling \code{twice}
with \code{addy} is not because the effect of \code{addy} is to
read $y$, not $x$.
\begin{lstlisting}
var x = 1;
var y = 2;
var twice = fun(f: func(int,int,[x]), y: int)[x]{
               var t = f(y); return f(t);
            };
var addx = fun(z:int)[x]{ return x + z; };
var addy = fun(z:int)[y]{ return y + z; };
var b = twice(addx, 3);                  /* OK */
var c = twice(addy, 3);               /* Error */
return b + c;
\end{lstlisting}

The solution to this problem is to parameterize functions with respect
to their effects~\cite{Tofte:1994uq,Pierce:2004fk}. In \funk{}, the
support for this comes from the \emph{effect abstraction}
$\alle{x}{f}$, which parameterizes its body $f$ with respect to
the variable $x$, and \emph{effect application} $\efapp{e}{y}$, in
which the result of $e$, an effect abstraction, is instantiated by
substituting $y$ for $x$. As an abbreviation, we write
$\alle{\vec{z}}{f}$ for $\alle{z_1}{\cdots \alle{z_n}{f}}$ and
$\efapp{e}{\vec{x}}$ for $\efapp{\efapp{e}{x_1}\cdots}{x_n}$.

To see this solution in action, we return to the previous example.
This time we parameterize \code{twice} with respect to $p$, a place
holder for a variable. The later uses of \code{twice} instantiate
$p$ with $x$ and $y$ respectively to get two different versions of
\code{twice} that work with \code{addx} and \code{addy}.
\begin{lstlisting}
...
var twice =
   <p> fun(f: func(int,int,[p]), y: int)[p] {
          var t = f(y); return f(t);
       };
...
var b = twice<x>(addx, 3);
var c = twice<y>(addy, 3);
return b + c;
\end{lstlisting}

Of course, we would like to parameterize with respect to arbitrary
numbers of variables, so that \code{twice} could be used with
functions with no effect or an effect with greater than one variable.
See \citet{Grossman:2002cr} for the generalization of effect
polymorphism that addresses this need.
From a syntactic standpoint, explicit effect application, as in
\code{twice<x>}, is rather heavyweight.  \citet{Grossman:2002cr} also
describe how to infer effect application, so we do not discuss this further here.

A programming language based on our design would provide syntactic
sugar for the definition of named, effect-polymorphic, recursive
functions. In this paper, we use the shorthand
\begin{lstlisting}
proc <$\vec{z}$> $f$($x$:$T_1$; $y$):$T_2$ [$\varphi$] { $s_1$ }
$s_2$
\end{lstlisting}
to mean
\begin{lstlisting}
var $f$ = (let $y$ = $y$ in fix $f$: $\allt{\vec{z}}{\funt{T_1}{\varphi}{T_2}}$. 
                      <$\vec{z}$> fun($x$:$T_1$)[$\varphi$]{ $s_1$ });
$s_2$
\end{lstlisting}
(The portions of the syntax related to $\vec{z}$, $x$,  $y$, and $\varphi$
are optional.)

\paragraph{Tail Calls and Effects}

To support space-efficient tail calls, \funk{} pops the current
procedure call frame \emph{prior} to performing the tail call.  To
make this safe, our type system does not allow tail calls to functions
that read from variables on the current frame.
We depart from the running example to show a classic
example~\cite{Appel:1992fk}, listed below, that was problematic for
region inference.  Ideally, this example uses $O(n)$ space; the
approach of \citet{Tofte:1994uq} used $O(n^2)$ but the later approach
of \citet{Aiken:1995fk} reduced the space to $O(n)$.
Indeed, the below example uses $O(n)$ in \funk{} because the two calls
in tail position are executed as tail calls.\footnote{For the purposes
  on this example, \funk{} is extended with support for \code{if}
  statements and lists in the obvious way.}

\begin{lstlisting}
proc s(i:int): int list {
   if (iszero(i)) { 
      return nil;
   } else {
      var p = dec(i); var sp = s(p);
      return cons(0, sp);
   }
}
proc f(n:int, x: int list):int [s]{
   var z = length(x);
   proc g(; f, n):int [s] {
      var s100 = s(100); return f(dec(n), s100);
   }
   if (iszero(n)) return 0;
   else return g();
}
var r = f(100,nil);
return r;
\end{lstlisting}

\noindent In \funk{}, the programmer knows for sure that every tail
call does not use extra stack space. Also, the programmer is notified
with an error message if a tail call could lead to a dangling
reference.


\paragraph{Take Care with Variables}

Particular care must be taken with the implementation of variables in
a language with a type and effect system such as \funk{}.
The straightforward approach leads to incorrect conclusions
during type checking. Consider the following program with two
variables with the name \code{x}.
\begin{lstlisting}
var x = 1;
proc f(a: int):int [x]{ return a + x; }
proc g(x: int):int [x]{ return f(x); }
var y = g(2);
return y;
\end{lstlisting}
Inside \code{g}, the tail call \code{return f(x)} has the effect
$\{x\}$.  A naive type checking algorithm would reject this tail call
because the parameter \code{x} is popped prior to the call. Of
course, that is the wrong \code{x}. 

A straightforward solution is for the compiler to rename all the
variables in the program to ensure each name is used in a variable
declaration only once. The result of such a renaming is shown below
for our example.
\begin{lstlisting}
var x$_1$ = 1;
proc f$_1$(a$_1$: int):int [x$_1$]{ return a$_1$ + x$_1$; }
proc g$_1$(x$_2$: int):int [x$_1$]{ return f$_1$(x$_2$); }
var y$_1$ = g$_1$(2);
return y$_1$;
\end{lstlisting}
Now the tail call \texttt{return f$_1$(x$_2$)} has the effect
\texttt{x$_1$}, which is benign because \texttt{x$_2$}, and not
\texttt{x$_1$}, is popped prior to the call. (Another solution is to
use the \citet{Bruijn:1972kx} representation for variables.)

\paragraph{Summary of the Design} 

The \funk{} calculus combines first-class functions with stack
allocation, enabling the full range of higher-order programming while
ensuring type safety and minimizing programmer-visible complexity
through a regionless type and effect system. Further, the design gives
the programmer control over efficiency by providing by-reference and
by-copy options for capturing the free variables of a function.

\section{Direct Semantics and Type Safety}
\label{sec:direct}

This section presents the dynamic and static semantics of \funk{}
and then a proof of type safety.

\subsection{Dynamic Semantics}

We give a substitution-based, single-step semantics for \funk{} that
takes inspiration from the semantics of \citet{Helsen:2000fk} and
\citet{Crary:1999fk}. However, to make stack allocation explicit, the
semantics operates on states of the form
\[
  \langle s, \kappa, \sigma, n \rangle 
\]
where $s$ is the body of the currently executing function, $\kappa$ is the
control stack, $\sigma$ is the value stack, and $n$ is the number of stack
values that are associated with the currently executing function.  On
a real machine, the control and value stacks are intertwined; our
formalization is more clear with them separated.  We write $\epsilon$ for
the empty stack and $v \cdot \sigma$ for pushing $v$ onto the front of the
stack $\sigma$. (So we represent stacks as cons-style lists.) We write
$|\sigma|$ for the length of stack $\sigma$ and concatenation of stacks is
juxtaposition, so the concatenation of $\sigma_1$ and $\sigma_2$ is written
$\sigma_1\sigma_2$.

In our semantics, variables are mapped to stack locations (represented
with natural numbers).  We index starting from the back of the stack
so that existing locations are not disturbed by growing the stack at
the front. Given the notation $\sigma_i$ for the normal front-to-back
indexing, we use the following notation for indexing back-to-front.
\[
  \sigma\{i\} = \sigma_{|\sigma| - 1 - i}
\]

As usual in a substitution-based semantics, the syntax of the language
must be slightly expanded for use by the dynamic semantics, which we
show in Figure~\ref{fig:internal-syntax} with the differences
highlighted in gray. The most important difference is that expressions
and effects include stack locations.  Stack locations have the form
$\ell_T$, combining the address $\ell$ with type $T$. This type annotation
is ignored by the dynamic semantics (see Section~\ref{sec:erasure});
it is merely a technical device used in the proof of type safety. The
need for these type annotations propagates to needing type annotations
in variable initialization and for function return types.  These type
annotations are straightforward to insert during the semantic analysis
(type checking) pass of a compiler.  One other difference is that we
add a top type $\top$ that is used by the type system to classify effect
parameters.  In Figure~\ref{fig:internal-syntax} we omit most of the
statements because they remain unchanged with respect to
Figure~\ref{fig:surface-syntax}.  The only change is adding a type
annotation on the variable initialization form.

\begin{figure}[tbp]
  \centering
\[
\begin{array}{lrcl}
  \text{stack index} & \ell & \in & \mathbb{N}\\
  \text{stack location} & l & ::= & \Gbox{x \mid \ell_T} \\
  \text{effects} & \varphi & ::= & l, \ldots, l \\
  \text{types} & T & ::= & \Gbox{\top} \mid \INT{} 
     \mid \funt{T}{\varphi}{T} \mid \allt{x}{T}\\
  \text{expressions} & e & ::= & \Gbox{l} \mid \efapp{e}{\Gbox{l}} \mid \ldots \\
  \text{abstractions} & f & ::= & \fune{x}{T}{\varphi}{\Gbox{T}}{s} \mid \ldots \\
  \text{statements} & s & ::= & \LET{x \Gbox{\of T}}{e}{s} \mid \ldots \\
  \\
  \text{values} & v & ::= & n 
                       \mid \fune{x}{T}{\varphi}{T}{s} \mid \alle{x}{e}  \\
  \text{value stack} & \sigma & ::= & \epsilon \mid v \cdot \sigma \\
  \text{frame} & F & ::= & (x\of T,s,n) \\
  \text{stack} & \kappa & ::= & \epsilon \mid F \cdot \kappa \\
   \text{state} & \varsigma & ::= & \langle s, \kappa, \sigma, n\rangle\\
  \text{observations} & o & ::= & n  \mid \kw{fun} \mid \kw{abs}  
\end{array}
\]
  \caption{Internal syntax and run-time structures.}
  \label{fig:internal-syntax}
\end{figure}

Figure~\ref{fig:eval} gives the evaluation of closed expressions
(expressions with no free variables) to values. Evaluation is
parameterized on the value stack $\sigma$ so that stack locations $\ell_T$
can be evaluated to their associated value. Most of the
equations in this definition are standard, such as the $\delta$ function
that gives meaning to all the primitive
operators~\cite{G.-D.-Plotkin:1975on}.
The evaluation of an effect application $\efapp{e}{\ell_T}$ first evaluates $e$
to an effect abstraction $\alle{x}{e'}$, substitutes $\ell_T$ for $x$ in
$e'$, and evaluates the result.

\begin{figure}[tbp]
\small
\boxed{\llbracket e \rrbracket_\sigma}
\begin{align*}
  \llbracket \ell_T \rrbracket_\sigma &= \sigma\{\ell \}  \\
  \llbracket n \rrbracket_\sigma &= n \\
  \llbracket \mathit{op}\kw{(}e_1,\ldots,e_n\kw{)} \rrbracket_\sigma & = \delta(\mathit{op}, (\llbracket e_1 \rrbracket_\sigma,\ldots,\llbracket e_n \rrbracket_\sigma)) \\
  \llbracket \fune{x}{T_1}{\varphi}{T_2}{s} \rrbracket_\sigma &= \fune{x}{T_1}{\varphi}{T_2}{s} \\
  \llbracket \LETE{x}{e_1}{e_2} \rrbracket_\sigma &= \llbracket [x {:=} \llbracket e_1 \rrbracket_\sigma] e_2 \rrbracket_\sigma \\
  \llbracket \kw{fix}\,x\of T.\,f \rrbracket_\sigma &= \llbracket [x{:=}\kw{fix}\,x\of T.\,f] f \rrbracket_\sigma \\
  \llbracket \alle{x}{f} \rrbracket_\sigma &= \alle{x}{f} \\
  \llbracket \efapp{e}{\ell_T} \rrbracket_\sigma &= \llbracket [x{:=}\ell_T]e' \rrbracket_\sigma \quad \text{if}\; \llbracket e\rrbracket_\sigma = \alle{x}{e'}
\end{align*}
  \caption{Evaluation of closed expressions.}
  \label{fig:eval}
\end{figure}

The single-step reduction relation over states is defined in
Figure~\ref{fig:reduce}. 
A variable initialization statement $\LET{x\of T}{e}{s}$ evaluates $e$
and pushes it on the value stack then substitutes its stack location
for $x$ in $s$.
A function call $\letcal{x}{e_1}{e_2}{s_1}$ evaluates $e_1$ to a
function value $\fune{y}{T_1}{\varphi}{T_2}{x}$, evaluates $e_2$ and pushes
it on the value stack, substitutes its stack location for $y$ in $s$,
then pushes the current call frame onto the control stack.
A tail call $\tailcal{e_1}{e_2}$ is similar except that it pops $n$
values from the stack.  (The \textit{drop} function returns a list
that lacks the first $n$ elements of the input list.)
The statement $\ret{e}$ evaluates $e$, pops $n$ values from the
stack, pushes the value of $e$ on the stack, and reinstates the top
frame from the control stack, substituting the stack location of $e$
for $x$ in $s$.
%

\begin{figure*}[tbp]
\boxed{\varsigma \longrightarrow \varsigma}
\begin{align}
 \langle (\LET{x\of T}{e}{s}), \kappa, \sigma , n \rangle & \longrightarrow \langle [x{:=}|\sigma|_{T}]s, \kappa, \llbracket e\rrbracket_\sigma \cdot \sigma, n+1\rangle 
 \tag{\textsc{Init}}\label{init}\\
 \langle (\letcal{x}{e_1}{e_2}{s_1}), \kappa, \sigma, n \rangle &
 \longrightarrow \langle [y{:=}|\sigma|_{T_1}]s_2, (x \of T_2, s_1, n) \cdot \kappa, \llbracket e_2\rrbracket_\sigma \cdot \sigma, 1 \rangle 
& \text{if } \llbracket e_1\rrbracket_\sigma = \fune{y}{T_1}{\varphi}{T_2}{s_2}
 \tag{\textsc{Call}}\label{call}\\
 \langle \tailcal{e_1}{e_2}, \kappa, \sigma, n \rangle &
 \longrightarrow \langle [y{:=}(|\sigma|{-}n)_{T_1}]s, \kappa, \llbracket e_2\rrbracket_\sigma \cdot \mathit{drop}(n,\sigma), 1 \rangle 
& \text{if } \llbracket e_1\rrbracket_\sigma = \fune{y}{T_1}{\varphi}{T_2}{s} 
 \tag{\textsc{TailCall}}\label{tail}\\
  \langle\ret{e}, (x\of T,s,n_2) \cdot \kappa, \sigma, n_1 \rangle 
& \longrightarrow \langle [x{:=}(|\sigma|{-}n_1)_T]s, \kappa, \llbracket e\rrbracket_\sigma \cdot \mathit{drop}(n_1,\sigma), n_2 + 1\rangle
 \tag{\textsc{Return}}\label{return}
\end{align}
  \caption{The single-step reduction relation.}
  \label{fig:reduce}
\end{figure*}

\begin{definition}
  The dynamic semantics of \funk{} is specified by the 
  partial function \textit{eval} defined by the following.
  \[
  \mathit{eval}(s) = \mathit{observe}(\llbracket e\rrbracket_\sigma) \text{ iff }
    \langle s, \epsilon, \epsilon, 0 \rangle \longrightarrow^{*} \langle \ret{e}, \epsilon, \sigma, n \rangle
  \]
  where $\mathit{observe}$ is defined on values as follows:
  \begin{align*}
    \mathit{observe}(n) &= n \\
    \mathit{observe}(\fune{x}{T_1}{\varphi}{T_2}{s}) &= \kw{fun} \\
    \mathit{observe}(\alle{x}{e}) &= \kw{abs}
  \end{align*}

\end{definition}

\subsection{Static Semantics}

As discussed in Section~\ref{sec:overview}, we require that all
variables be uniquely named in a pre-processing pass of a compiler.

The type system for \funk{} is inductively defined in
Figure~\ref{fig:type-system}.  The judgment for well-typed expressions
has the form $\Gamma; \varphi_1 \vdash e : T$ while the judgment for well-typed
statements has the form $\Gamma; \varphi_1; \varphi_2 \vdash s : T$. The $\Gamma$ is a type
environment (symbol table), mapping variables to types, and is
described in more detail below.  The effect $\varphi_1$ specifies which
variables may be read from; the second effect $\varphi_2$ in the context
for statements keeps track of which variables are popped from the
value stack upon a return from the current call frame, that is, it
keeps track of the parameters and local variables. We discuss the
important aspects of the type system in the following paragraphs after
introducing our slightly non-standard type environments.

\paragraph{Type Environments}
We write $\emptyset$ for the empty type environment and $\Gamma,x\of T$ for
extending the type environment $\Gamma$ with the binding of a
stack-allocated variable $x$ to type $T$. We write $\Gamma,x\of
\mathsf{copy}\,T$ for extending the type environment with 
variables that are bound by the \code{let} or \code{fix} expressions.
The \textsf{copy} annotation helps the type system distinguish between
reads from stack-allocated variables, which count as an effect, and
reads from \code{let}-bound or \code{fix}-bound variables, which do not.
Looking up a variable $y$ in an environment $\Gamma$, written $\Gamma(y)$, is
defined as follows.
\begin{align*}
  (\Gamma,x\of T)(y) = 
  \begin{cases}
    T & \text{if } x = y \\
    \Gamma(y) & \text{otherwise}
  \end{cases} \\
  (\Gamma,x\of \mathsf{copy}\, T)(y) = 
  \begin{cases}
    \mathsf{copy}\,T & \text{if } x = y \\
    \Gamma(y) & \text{otherwise}
  \end{cases}
\end{align*}

\paragraph{Well-typed Expressions}

An occurrence of a stack-allocated variable must both be in scope and
in the declared effect of the surrounding function. In contrast, a
\code{let}-bound or \code{fix}-bound variable need only be in scope.
During execution, stack locations are substituted for
stack-allocated variables. A stack location is well typed if it is in the
declared effect. This rule is simple but
subtle. A well-typed program in mid execution may have functions that
contain dangling stack locations. The type system ensures that such
functions are never called. Thus, the typing rule for stack locations
does not look at the actual value stack, but instead merely checks
that the stack location is in the declared effect of the surrounding
function.

A function expression is well-typed if its body is well typed in the
declared effect $\varphi_2$ enlarged with parameter $x$. The second effect
context is $\{ x \}$ because $x$ is popped from the stack upon return
from this function.
The typing rules for the rest of the expressions are standard.

\paragraph{Well-typed Statements}

The typing rule for variable initialization is straightforward, given
that it allocates the variable on the stack.  The role of the effect
annotations in function types can be seen in the typing rules for
function call and tail call.  In both cases, the effect of the
function $\varphi_3$ must be contained in the current effect $\varphi_1$. In
addition, for tail calls, $\varphi_3$ must not overlap with $\varphi_2$, the
local variables of the current function.

The rules for tail call and return must both prevent the escape of
upward funargs with dangling references. The rules accomplish this by
requiring that the free variables in the returned value's type not
include any local variables. The free variables of a type, written
$\mathrm{fv}(T)$, is defined as follows.
\begin{align*}
  \mathrm{fv}(\INT) &= \emptyset \\
  \mathrm{fv}(\funt{T_1}{\varphi}{T_2}) &= \mathrm{fv}(T_1) \cup \varphi \cup \mathrm{fv}(T_2)\\
  \mathrm{fv}(\allt{x}{T}) &= \mathrm{fv}(T) - \{ x \} 
\end{align*}


\begin{figure}[tbp]
\boxed{\Gamma \vdash \varphi}
\[
   \Gamma \vdash \varphi \text{ iff } \forall x \in \varphi.\, \exists T.\, \Gamma(x) = T
\]
\boxed{\Gamma \vdash T}
\begin{gather*}
  \inference{}{\Gamma \vdash \top}
  \inference{}{\Gamma \vdash \INT}
  \inference{
    \Gamma \vdash T_1 & \Gamma \vdash T_2 \\ \Gamma \vdash \varphi 
  }{
    \Gamma \vdash \funt{T_1}{\varphi}{T_2}
  }
  \inference{\Gamma,x\of \top \vdash T }
       {\Gamma \vdash \allt{x}{T}}
\end{gather*}
\boxed{\Gamma; \varphi \vdash e : T}
\begin{gather*}
\inference{\Gamma(x) = T& x \in \varphi}{\Gamma; \varphi \vdash x : T}
\inference{\Gamma(x) = \mathsf{copy}\,T}{\Gamma; \varphi \vdash x : T}
\inference{
  \ell_T \in \varphi
  }
  {\Gamma; \varphi \vdash \ell_T : T}
\\[1ex]
\inference{}{\Gamma; \varphi \vdash n : \INT}
\;
\inference{\mathit{typeof}(\mathit{op}) {=} (T_1,\ldots,T_n, T_r) \\ 
     \Gamma; \varphi \vdash e_i : T_i \text{ for } i\in \{ 1,\ldots,n\} }
    {\Gamma; \varphi \vdash \mathit{op}(e_1,\ldots,e_n) : T_r}
\\[1ex]
\inference{
  \Gamma \vdash \funt{T_1}{\varphi_2}{T_2} &
  \Gamma,x\of T_1; \varphi_2 \cup \{ x \}; \{ x \} \vdash s : T_2
}{
\Gamma; \varphi_1 \vdash \fune{x}{T_1}{\varphi_2}{T_2}{s} : \funt{T_1}{\varphi_2}{T_2}
}
\\[1ex]
\inference{
  \Gamma; \varphi \vdash e_1 : T_1 \\
 \Gamma,x\of \mathsf{copy}\,T_1; \varphi \vdash e_2 : T_2
}{
 \Gamma; \varphi \vdash \LETE{x}{e_1}{e_2} : T_2
}
\quad
\inference{
  \Gamma,x\of \mathsf{copy}\,T; \varphi \vdash f : T
}{
  \Gamma; \varphi \vdash \kw{fix}\,x\of T.\,f : T
}
\\[1ex]
\inference{
  \Gamma,x:\top; \emptyset \vdash e : T
}{
  \Gamma; \varphi \vdash \alle{x}{e} : \allt{x}{T}
}
\quad
\inference{
  \Gamma; \varphi \vdash e : \allt{x}{T}
}{
  \Gamma; \varphi \vdash \efapp{e}{l} : [x{:=}l]T
}
\end{gather*}
\boxed{\Gamma; \varphi ; \varphi \vdash s : T}
\begin{gather*}
\inference{
  \Gamma ; \varphi_1 \vdash e : T_1 \\
  \Gamma,x\of T_1; \varphi_1 \cup \{x\}; \varphi_2 \cup \{ x \} \vdash s : T_2
}{
  \Gamma; \varphi_1; \varphi_2 \vdash \LET{x\of T_1}{e}{s} : T_2
}
\\[1ex]
\inference{
    \Gamma; \varphi_1 \vdash e_1 : \funt{T_1}{\varphi_3}{T_2} 
  & \Gamma; \varphi_1 \vdash e_2 : T_1
  \\ \Gamma,x\of T_2; \varphi_1 \cup \{ x \}; \varphi_2 \cup \{ x \} \vdash s : T_3
  & \varphi_3 \subseteq \varphi_1 
}
{
\Gamma; \varphi_1; \varphi_2 \vdash \letcal{x}{e_1}{e_2}{s} : T_3
}
\\[1ex]
\inference
{
  \Gamma; \varphi_1 \vdash e_1 : \funt{T_1}{\varphi_3}{T_2} 
  & \Gamma; \varphi_1 \vdash e_2 : T_1
  \\ \varphi_3 \subseteq \varphi_1 - \varphi_2
  & \mathrm{fv}(T_2) \cap \varphi_2 = \emptyset 
}
{
\Gamma; \varphi_1; \varphi_2 \vdash \tailcal{e_1}{e_2} : T_2
}
\\[1ex]
\inference
{
  \Gamma; \varphi_1 \vdash e : T
  & \mathrm{fv}(T) \cap \varphi_2 = \emptyset 
}
{
\Gamma; \varphi_1; \varphi_2 \vdash \ret{e} : T
}
\end{gather*}
  \caption{The typing rules for expressions and statements.}
  \label{fig:type-system}
\end{figure}

\begin{figure}[tbp]
\boxed{\vdash \Gamma}
\begin{gather*}
  \inference{}{\vdash \emptyset}
  \quad
  \inference{\vdash \Gamma & \Gamma \vdash T}{\vdash \Gamma,x\of T}
\end{gather*}
\boxed{\vdash \sigma : \Sigma}
\begin{gather*}
  \inference{}{\vdash \epsilon : \epsilon} 
  \quad
  \inference{\emptyset ; \emptyset \vdash v : T & \vdash \sigma : \Sigma}{\vdash v \cdot \sigma : T \cdot \Sigma}
\end{gather*}
\boxed{\Sigma \vdash \varphi}
\vspace{-15pt}
\[
   \Sigma \vdash \varphi \equiv \forall \ell_T \in \varphi.\, \Sigma\{ \ell \}  = T
\]
\boxed{\Sigma \vdash \kappa : T \Rightarrow T}
\begin{gather*}
  \inference{}{\Sigma \vdash \epsilon : T \Rightarrow T}
  \quad
  \inference{
    \Sigma \vdash \varphi_1
  & \mathit{drop}(n,\Sigma) \vdash \varphi_1 - \varphi_2
  \\ \emptyset,x\of T_1; \varphi_1 \cup \{ x \} ; \varphi_2 \cup \{ x \} \vdash s : T_2
  \\ \emptyset \vdash T_1 
  & \mathit{drop}(n,\Sigma) \vdash \kappa : T_2 \Rightarrow T_3}
   {\Sigma \vdash (x\of T_1,s,n) \cdot \kappa : T_1 \Rightarrow T_3}
\end{gather*}

\boxed{\vdash \varsigma : T}
\[
\inference{
  \Sigma \vdash \varphi_1
& \mathit{drop}(n,\Sigma) \vdash \varphi_1 - \varphi_2
& \emptyset ; \varphi_1; \varphi_2 \vdash s : T_1
\\ \vdash \sigma : \Sigma
& \mathit{drop}(n,\Sigma) \vdash \kappa : T_1 \Rightarrow T_2
}
{\vdash \langle s, \kappa, \sigma, n\rangle : T_2}
\]
\boxed{\vdash o : T}
\begin{gather*}
\inference{}{\vdash n : \INT}\;
\inference{}{\vdash \kw{fun} : \funt{T_1}{}{T_2}}\;
\inference{}{\vdash \kw{abs} : \allt{x}{T}}
\end{gather*}
  \caption{The typing rules for states and observations.}
  \label{fig:type-system-state}
\end{figure}

\subsection{Substitution vs. Environments}

We originally formulated the dynamic semantics of \funk{} using
environments instead of substitution. The authors prefer environments
because they bring the semantics a step closer to a real
implementation and because they more closely correspond to a
programmer's view of variables. However, the definition of well-typed
states became much more complex because it was difficult to define a
notion of well-typed environment that interacted properly with effect
polymorphism. Switching to a substitution-based semantics removed the
need for well-typed environments, which simplified the definition of
well-typed states and the proof of type safety.

\subsection{Type Safety}
\label{sec:type-safety}

Type safety for \funk{} means that a well-typed program cannot get
stuck and that the result of the program matches its static type. The
notion of \emph{stuck} models untrapped errors such as memory
errors~\cite{Cardelli:1997fk}.  Technically, getting stuck means
getting into a state that neither has a subsequent state (as defined
by the step relation $\longrightarrow$) nor is a final state,
as defined below.

\begin{definition}[Final State]
  $\mathit{final}(\varsigma)$ if and only if the state $\varsigma$
  is of the form $\langle \ret{e}, \epsilon, \sigma, n\rangle$.
  Given a final state $\varsigma$, we write $\mathit{result}(\varsigma)$
  for $\mathit{observe}(\llbracket e\rrbracket_\sigma)$.
\end{definition}

The formal statement of type safety unravels the negative ``not
stuck'' to arrive at the following positive form. The definition of a
well-typed observation (for $\vdash \mathit{result}(\varsigma) : T$
below) is given in Figure~\ref{fig:type-system-state}.

\begin{theorem}[Type Safety]
  \label{thm:type-safety}
  If $\emptyset; \emptyset \vdash s : T$ and $\langle s, \epsilon, \epsilon, 0\rangle \longrightarrow^{*} \varsigma$, then either
  $\mathit{final}(\varsigma)$ and $\vdash \mathit{result}(\varsigma) : T$ or $\varsigma \longrightarrow \varsigma'$
  for some $\varsigma'$.
\end{theorem}

In the following, we give a proof for the Type Safety Theorem,
including the statement but not the proofs of the major lemmas. (This
proof is related to the proof of type safety by \citet{Helsen:2000fk}
for the region calculus in that we take the syntactic approach and
have a substitution-based semantics.)
The proofs of the lemmas are in the accompanying technical report.
(See the supplemental material.)

As usual, the proof of type safety proceeds by induction on the
reduction sequence $\langle s, \epsilon, \epsilon, 0\rangle \longrightarrow^{*} \varsigma$. The induction step
of the proof concerns a single reduction, from some intermediate state
$\varsigma_1$ to the next state $\varsigma_2$. What needs to be proved for such
intermediate states is that a well-typed state must either be a final
state or it can take a step to another well-typed state (progress and
preservation).

\begin{lemma}[Step Safety]
  \label{lem:step-safety}
  If $\vdash \varsigma_1 : T$, then
  either $\mathit{final}(\varsigma_1)$ or
  $\varsigma_1 \longrightarrow \varsigma_2$ and $\vdash \varsigma_2 : T$ for some $\varsigma_2$.
\end{lemma}

\noindent The Step Safety proof hinges on the definition of well-typed states.

\paragraph{Well-typed States}

The definition of a well-typed state is given in
Figure~\ref{fig:type-system-state}. Getting this definition right,
such that we can prove Step Safety, is the main challenge in the
overall proof of Type Safety. The definition of well-typed states
relies on several auxiliary judgments which roughly correspond to the
ones used by \citet{Crary:1999fk} and \citet{sabry02:_minml}.  The
judgment $\vdash \sigma : \Sigma$ is for well-typed value stacks, where $\Sigma$ is a
list of types.  The judgment $\Sigma \vdash \varphi$ says that $\Sigma$
\textit{satisfies} $\varphi$, which is to say, any value stack of type $\Sigma$
can be used to safely execute a statement that requires effect $\varphi$.
The judgment $\Sigma \vdash \kappa : T_1 \Rightarrow T_2$ is for well-typed control stacks.
The control stack $\kappa$ is expecting a value of type $T_1$ (returned
from the current call frame) and ultimately produces a value of
type $T_2$, so long as the value stack has type $\Sigma$.

The rule for a well-typed state $\langle s, \kappa, \sigma, n \rangle$ requires that the
statement $s$ be well-typed in the context of an effect declaration
$\varphi_1$ and local variables $\varphi_2$. The value stack must be well typed
($\vdash \sigma : \Sigma$) and, critically, its type $\Sigma$ must satisfy the effect
$\varphi_1$ which is needed to safely execute statement $s$. Further, popping $n$
items from $\Sigma$ must yield a stack typing that satisfies $\varphi_1 -
\varphi_2$, that is, the current live variables but not including the $n$ local
variables to be popped. The typing rule for extending the control stack
with a call frame is analogous to the typing rule for states, the only
difference being that a call frame is expecting a return value of type
$T_1$ that it binds to the stack-allocated variable $x$.

\subsubsection{Expression Evaluation Safety}

All of the reduction rules (Figure~\ref{fig:reduce}) require
evaluating an expression, which means we need a notion of type safety
for expressions. This is captured in the Evaluation Safety Lemma,
stated below. We only deal with closed expression ($\Gamma=\emptyset$) because,
in a well-typed program, the variables are substituted away before we
come to evaluate an expression.  The process of evaluation replaces
stack locations with their associated values, so the resulting value
is well-typed in an empty effect context. There may, however, still be
effects in the type $T$ (if the value is a function and the body of
the function contains stack locations). The effects in $T$ dictate how
the resulting value can be used. To ensure that the stack locations
correspond to values of the appropriate type, this lemma includes the
premises $\vdash \sigma : \Sigma$ and $\Sigma \vdash \varphi$.

\begin{lemma}[Evaluation Safety]\label{lem:eval-safety}
  If $\emptyset \vdash \varphi$, $\emptyset; \varphi \vdash e : T$, $\vdash \sigma : \Sigma$, and $\Sigma \vdash \varphi$, then
  $\emptyset; \emptyset \vdash \llbracket e \rrbracket_\sigma : T$.
\end{lemma}

Looking at the various cases for evaluation in Figure~\ref{fig:eval},
there are several that require knowing that a subexpression evaluates
to a particular form of value. For example, to evaluate an effect
application $\efapp{e}{\ell_T}$, the expression $e$ must evaluate to an
effect application, which has the form $\alle{x}{e'}$. We repeat
this evaluation rule below.
\[
  \llbracket \efapp{e}{\ell_T}] \rrbracket_\sigma = \llbracket [x{:=}\ell_T]e' \rrbracket_\sigma \quad \text{if}\; \llbracket e\rrbracket_\sigma = \alle{x}{e'}
\]
From the typing rule for $\efapp{e}{\ell_T}$, we see that $e$ has type $\allt{x}{T'}$ for some $T'$, and by the induction hypothesis, the value $\llbracket
e \rrbracket_\sigma$ also has type $\allt{x}{T'}$. The Canonical Forms Lemma, stated
below, tells us that the value $\llbracket e \rrbracket_\sigma$ must therefore be an
effect abstraction (and similar facts about the other types).
\begin{lemma}[Canonical Forms]\label{lem:canonical}
  If $\emptyset; \emptyset \vdash v : T$, then exactly one of the following holds.
  \begin{enumerate}
  \item $T = \INT$ and $v = n$ for some $n$, or
  \item $T = \funt{T_1}{\varphi}{T_2}$ and $v = \fune{x}{T_1}{\varphi}{T_2}{s}$
    for some $x$,$T_1$,$T_2$,$\varphi$,$s$, or
  \item $T = \allt{x}{T'}$ and $v = \alle{x}{e}$ for some $x$ and $e$.
  \end{enumerate}
\end{lemma}

The evaluation rules for \code{let} and \code{fix} involve
substituting a (closed) expression for a variable.
Thus, we need to know that substitution, given well-typed expressions,
produces a well-typed expression. Because expressions contain
statements, and vice versa, this Lemma must be proved simultaneously
for expressions and statements.
\begin{lemma}[Substitution Preserves Types] \ \\
  \label{lem:subst}
  Suppose that $\emptyset; \emptyset \vdash e : T_1$ and $\vdash \Gamma_1,x\of T_1,\Gamma_2$.
  \begin{enumerate}
  \item If $\Gamma_1,x\of T_1,\Gamma_2; \varphi_1 \vdash e_1 : T_2$, then
    $\Gamma_1,\Gamma_2; \varphi_1 \vdash \subs{x}{e}{e_1} : T_2$.
  \item If $\Gamma_1,x\of T_1,\Gamma_2; \varphi_1; \varphi_2 \vdash s : T_2$, \\
   then $\Gamma_1,\Gamma_2; \varphi_1; \varphi_2 \vdash \subs{x}{e}{s} : T_2$.
  \end{enumerate}
\end{lemma}
Such a substitution lemma often takes a different form, with no
$\Gamma_2$, and instead relies on a Permutation Lemma to keep the binding
for $x$ on the right-most end of the type environment while going
under a function expression. However, we can not permute environments
in \funk{} because variables may appear in types, so the particular
sequencing of variables in the environment is rather important.  (For
similar reasons, permutation does not hold in polymorphic calculi
such as System F.) The expression $e$ being substituted-in may be
spliced into a context with a different typing environment with
possibly many variables in scope. This is irrelevant to the typing of
$e$ because $e$ does not have any free variables. The following
Environment Weakening Lemma makes this precise.

\begin{lemma}[Environment Weakening] \ \\
 \label{lem:weakening}  
 Suppose $\vdash \Gamma$, $\emptyset \vdash \varphi_1$, and $\emptyset \vdash \varphi_2$.
  \begin{enumerate}
  \item If $\emptyset; \varphi_1 \vdash e : T$, then $\Gamma; \varphi_1 \vdash e : T$.
  \item If $\emptyset; \varphi_1; \varphi_2 \vdash s : T$, then $\Gamma; \varphi_1; \varphi_2 \vdash s : T$.
  \end{enumerate}
\end{lemma}

The expression $e$ being substituted also goes from a context with no
effects to a context with possibly many effects. Thus, we also need
weakening for effects. The following Effect Weakening Lemma does not
need to be proved simultaneously for statements because the typing of
the body of a function does not depend on the current effect.

\begin{lemma}[Effect Weakening]
  \label{lem:effect-weakening}  
  If $\Gamma; \varphi \vdash e : T$, $\varphi \subseteq \varphi'$, and $\Gamma \vdash \varphi'$, then $\Gamma; \varphi' \vdash e : T$.
\end{lemma}

The evaluation rule for effect application involves substituting a
stack location for a variable. 
Substituting stack locations is more involved than substituting
expressions because stack locations occur in effects and therefore in
types. The proof that substitution of stack locations preserves types
requires simultaneously proving this property for types, expressions,
and statements.

\begin{lemma}[Substitution of Stack Locations Preserves Types] \ \\
  \label{lem:subst-loc}
  Let $[X]$ stand for $\subs{x}{\ell_{T_1}}{X}$ in the following.\\
  Suppose $\Gamma_1,x\of T_1,\Gamma_2 \vdash \varphi_1$ and $\Gamma_1,x\of T_1,\Gamma_2 \vdash \varphi_2$.
  \begin{enumerate}
  \item If $\Gamma_1,x\of T_1,\Gamma_2 \vdash T_2$, then $\Gamma_1,[\Gamma_2] \vdash [T_2]$.
  \item If $\Gamma_1,x\of T_1,\Gamma_2; \varphi_1 \vdash e : T_2$, then
    $\Gamma_1,[\Gamma_2]; [\varphi_1] \vdash [e] : [T_2]$.
  \item If $\Gamma_1,x\of T_1,\Gamma_2; \varphi_1; \varphi_2 \vdash s : T_2$, \\
    then $\Gamma_1,[\Gamma_2]; [\varphi_1]; [\varphi_2] \vdash [s] : [T_2]$.
  \end{enumerate}
\end{lemma}

Of course, to handle the case of evaluating a primitive operator, we
require that all the primitive operators be type safe.
\begin{constraint}[Type safe primitive operators]
  \label{constraint:primop}
  If $\mathit{typeof}(\mathit{op}) = (T_1,\ldots,T_n,T_r)$ and $\emptyset ;\emptyset \vdash
  v_i : T_i$ for $i \in \{ 1,\ldots, n\}$, then $\delta(\mathit{op}, (v_1,\ldots,v_n)) = v'$ 
  and $\emptyset ;\emptyset \vdash v' : T_r$ for some $v'$.
\end{constraint}

\noindent No other lemmas are required to prove Evaluation Safety.

\subsubsection{Step Safety}

The proof of Step Safety requires a few more key lemmas.
Lemma~\ref{lem:cap-stmt-return} is a consequence of how the type
system is designed to prevent upward funargs, that is, the return of 
functions with dangling references to local variables.

\begin{lemma}[Locals not in Return Type]
  \label{lem:cap-stmt-return}
  Suppose $\Gamma \vdash \varphi_1$ and $\Gamma \vdash \varphi_2$.  If $\Gamma; \varphi_1; \varphi_2 \vdash s :
  T$, then $\mathrm{fv}(T) \cap \varphi_2 = \emptyset$.
\end{lemma}

\noindent The substitution of stack addresses into the return type of
a function doesn't change the return type (which would break type
preservation), because of the above lemma together with the following.

\begin{lemma}[Substitution of Non-free Variables]
  \label{lem:cap-id}
  If $x \notin \mathrm{fv}(T)$, then $\subs{x}{\ell_{T'}}{T} = T$.
\end{lemma}

\noindent A function may have fewer effects than are present in the
calling context, so we need the following weakening lemma.

\begin{lemma}[Satisfaction Weakening]
  \label{lem:capability-weaken}
 If $\Sigma \vdash \varphi_1$ and $\varphi_2 \subseteq \varphi_1$, then $\Sigma \vdash \varphi_2$.
\end{lemma}

\noindent Finally, in tail calls and when returning from a function,
we pop the values corresponding to parameters and local variables from
the stack, but the remaining stack is still well typed.

\begin{lemma}[Drop Stack]
  \label{lem:drop-stack}  
  If $\vdash \sigma : \Sigma$, then $\vdash \mathit{drop}(n,\sigma) : \mathit{drop}(n,\Sigma)$.
\end{lemma}

\noindent The proof of Step Safety (Lemma~\ref{lem:step-safety}) then
proceeds by case analysis on the statement component of the state
$\varsigma_1$.

\section{Erasure-based Semantics}
\label{sec:erasure}

\begin{figure}[tbp]
  \centering
\[
\begin{array}{rcl}
  \erase{x} &=&x \\[1ex]
  \erase{n} &=&n \\
  \erase{\mathit{op}(e_1\ldots e_n)} &=&\mathit{op}(\erase{e_1}\ldots\erase{e_n}) \\
  \erase{\LETE{x}{e_1}{e_2}} &=&\LETE{x}{\erase{e_1}}{\erase{e_2}} \\
  \erase{\mathtt{fix}\, x\of T.\ f} &=&\mathtt{fix}\ x.\ \erase{f} \\
  \erase{\efapp{e}{l}} &=&\erase{e} \\[1ex]
  \erase{\fune{x}{T}{\varphi}{T}{s}} &=&\kw{fun(}x\kw{)}\{\erase{s}\} \\
  \erase{\alle{x}{f}} &=&\erase{f} \\[1ex]
  \erase{\LET{x\of T}{e}{s}} &=&\LET{x}{\erase{e}}{\erase{s}}\\
  \erase{\letcal{x}{e_1}{e_2}{s}}&=&\letcal{x}{\erase{e_1}}{\erase{e_2}}{\erase{s}}\\
  \erase{\ret{e_1(e_2)}}&=&\ret{\erase{e_1}(\erase{e_2})}\\
  \erase{\ret{e}}&=&\ret{\erase{e}}
\end{array}
\]  
\caption{Erasure function}
  \label{fig:erasure}
\end{figure}

An important question from both the semantic and implementation
perspective is whether the execution of an effect abstraction depends
in any way on the effects with which it is instantiated.  We answer
this question in the negative by demonstrating that \funk{} can be
implemented using an erasure-based approach.  A more direct argument
for this property, known as relational parametricity, can be
accomplished via a proof based on logical relations, but we leave that
for future work.

Concretely, because effect abstractions do not depend on their
arguments, it is unnecessary to perform effect substitutions at
runtime. In the direct semantics of \funk{}, evaluation of effect
applications $\efapp{e}{\ell_T}$ involves substitution of the stack
location $\ell_T$ into the subterms of $e$.  However, since effect
abstraction variables may be substituted by concrete variables of any
type, effect variables may only be used in effect applications and
function effect annotations, which do not interact with the rest of
the program at runtime --- in other words, \funk{} programs are
parametric with regard to effect polymorphism.  

We remove both the type and effect annotations by an erasure function,
defined in Figure \ref{fig:erasure}. We even erase effect abstractions
and effect applications. The syntactic restriction in \funk{} that the
body of an effect abstraction is an abstraction (a value) means that
there is no need to delay the evaluation of the body (because the body
is already a value).


We may evaluate erased programs with similar rules to those which
evaluate \funk{} and we refer to the resulting evaluation function as
$\mathit{eval}_{\mathit{erased}}$.
The similarity of these rules to those for \funk{} leads us to the
following:
\begin{proposition}[Preservation of semantics under erasure]
  If $\emptyset;\emptyset;\emptyset\vdash s : T$, then 
  $\mathit{eval}(s)= \mathit{eval}_{\mathit{erased}}(\erase{s})$.
\end{proposition}
We have performed rigorous testing of this property and found no
violations. It should be straightforward to prove this preservation of
semantics via a simulation argument.


\section{Translation to the Region Calculus}
\label{sec:translate-to-regions}

In this section we relate \funk{} to a region calculus, in particular,
a calculus similar to Henglein's ETL and RTL
calculi~\cite{Pierce:2004fk}.  The goal of this section is not to
provide an aid to implementation, but to aid the reader in
understanding the relation between our work and prior work on regions.
Thus, we choose a relatively prototypical region calculus at the cost
of losing efficient tail calls. (There are other region-based calculi
that would enable efficient tail
calls~\cite{Aiken:1995fk,Crary:1999fk}.)

The syntax for the region calculus is defined in
Figure~\ref{fig:regions}. The type system is
standard~\cite{Pierce:2004fk}.  We use the variable $r$ to range over
expressions, to easily distinguish region calculus expressions from
\funk{} expressions.  The form $\kw{new}\,\rho.\,r$ allocates an
empty region and binds it to the name $\rho$ for use in $r$.  The form
$r\,\kw{at}\,\rho$ allocates a location in region $\rho$, evaluates
expression $r$ and stores its value into that location, then returns
the location. The form $r\,!\,\rho$ evaluates $r$ to a location and
dereferences the location, that is, extracts the value stored at that
location of region $\rho$.  In addition to the forms in
Figure~\ref{fig:regions}, we use the following syntactic sugar:
$\LETE{x}{r_1}{r_2} \triangleq (\lambda x. r_2)\, r_1 $.

\begin{figure}[tbp]
\[
\begin{array}{lrcl}
  \text{region variables}        & \rho \\
  \text{effects}        & \varphi & ::= & \{ \rho, \ldots , \rho \} \\
  \text{types} & T & ::= & \INT{} \mid T\overset{\varphi}{\to}T \mid \forall \rho.\, T \mid T\,\kw{at}\,\rho\\
  \text{expressions}   & r & ::= & 
     x \mid n \mid \mathit{op}(r,\ldots,r) \\
  &&&
     \mid \lambda x.\, r
     \mid r\, r
     \mid \kw{fix}\,x.\, r \\
  &&&
     \mid \kw{new}\,\rho.\, r \mid r\,\kw{at}\, \rho \mid r\,!\, \rho\\
  &&&
    \mid \Lambda\rho.\, r \mid \efappr{r}{\rho}
\end{array}
\]
  \caption{A region calculus}
  \label{fig:regions}
\end{figure}

Figure~\ref{fig:translate-to-regions} shows our translation from
\funk{} to the region calculus. The translation functions take two
parameters, the first some syntax (expressions, etc.) and the second a
partial function from variables to regions.  The main idea behind the
translation is that we allocate a new region for each variable that is
stack allocated in \funk. This idea is realized in the translation for
initialization statements shown below.  The $\kw{new}\,\rho$
allocates the region for the variable $x$ and the expression $\toreg{
  e }{R}\,\kw{at}\,\rho$ assigns the value of $e$ into a new
location in the region. This location is bound to $x$. The following
statement $s$ is translated using the mapping $R(x{:=}\rho)$, that is,
extending $R$ with the association of region $\rho$ to $x$.
\begin{align*}
  \toreg{ \LET{x}{e}{s} }{R} &= 
  \kw{new}\,\rho.\, 
    \LETE{x}{\toreg{ e }{R}\,\kw{at}\,\rho}{\toreg{s}{R(x{:=}\rho)}} \\
    & \text{where } \rho \text{ is fresh}
\end{align*}
The second half of the story is the translation of variable occurrences.
If $R$ associates a region $\rho$ with the variable $x$, then we
translate the variable $x$ to $x\,!\,\rho$, which dereferences the location
that is bound to $x$. Otherwise, the variable $x$ must have been bound
by either a \code{let} or \code{fix}, in which case there is no need
to dereference.
\begin{align*}
  \toreg{x}{R} &=
  \begin{cases}
   x\,!\,\rho & \text{if } R(x) = \rho \\
   x  & \text{otherwise}
  \end{cases} 
\end{align*}

\begin{figure}[tbp]
\boxed{\toreg{T}{R}}
\vspace{-18pt}
\begin{align*}
  \llbracket \INT \rrbracket_R &= \INT \\
  \llbracket \funt{T_1}{\varphi}{T_2} \rrbracket_R &=  \llbracket T_1 \rrbracket_R\overset{\toreg{\varphi}{R}}{\to}\llbracket T_2 \rrbracket_R \\
  \llbracket \allt{x}{T} \rrbracket_R &= \forall \rho.\, \llbracket T\rrbracket_{R(x{:=}\rho)}
\end{align*}

\boxed{\toreg{e}{R}}
\vspace{-15pt}
\begin{align*}
  \toreg{x}{R} &=
  \begin{cases}
   x\,!\,\rho & \text{if } R(x) = \rho \\
   x  & \text{otherwise}
  \end{cases} \\
  \toreg{n}{R} &= n \\
  \toreg{ \mathit{op}(e_1,\ldots,e_n) }{R} &= 
      \mathit{op}(\toreg{ r_1 }{R},\ldots,\toreg{ r_n }{R}) \\
  \toreg{ \fune{x}{T_1}{\varphi}{T_2}{s} }{R} &= 
      \lambda y.\, \kw{new}\,\rho.\, \LETE{x}{y\,\kw{at}\,\rho}{\toreg{s}{R(x{:=}\rho)}} \\
  & \text{where } y,\rho \text{ are fresh} \\
  \toreg{ \alle{x}{f} }{R} &= 
     \Lambda \rho.\, \toreg{f}{R(x{:=}\rho)} \\
  & \text{where } \rho \text{ is fresh} \\
  \toreg{ \LETE{x}{e_1}{e_2} }{R} &= 
      \LETE{x}{\toreg{ e_1 }{R}}{\toreg{ e_2 }{R}}  \\
 \toreg{ \kw{fix}\,x\of T.\, f }{R} &= 
    \kw{fix}\,x.\, \toreg{f}{R} \\
  \toreg{ \efapp{e}{x} }{R} &= \efappr{\toreg{e}{R}}{R(x)} 
\end{align*}

\boxed{\toreg{s}{R}}
\begin{align*}
  \toreg{ \LET{x}{e}{s} }{R} &= 
  \kw{new}\,\rho.\, \LETE{x}{\toreg{ e }{R}\,\kw{at}\,\rho}{\toreg{s}{R(x{:=}\rho)}} \\
  & \text{where } \rho \text{ is fresh} \\
  \toreg{ \letcal{x}{e_1}{e_2}{s} }{R} &=
    \kw{new}\,\rho.\, \LETE{x}{e}{\toreg{s}{R(x{:=}\rho)}} \\
  & \text{where } \rho \text{ is fresh} \text{ and } e = (\toreg{e_1}{R}\,\toreg{e_2}{R})\,\kw{at}\,\rho \\
  \toreg{ \tailcal{e_1}{e_2} }{R} &= \toreg{e_1}{R}\,\toreg{e_2}{R}\\
  \toreg{\ret{e}}{R} &= \toreg{e}{R}
\end{align*}

\boxed{\relate{\Gamma}{R}{\Gamma_r}}
\begin{gather*}
  \inference{}{ \relate{\emptyset}{R}{\emptyset} } \qquad
  \inference{
    T \neq \top & \relate{\Gamma}{R}{\Gamma_r}
  }{
    \relate{(\Gamma,x\of T)}{R}{(\Gamma_r,y\of \llbracket T\rrbracket_R,x\of \llbracket T\rrbracket_R\,\kw{at}\,R(x))}
  }
  \\[2ex]
  \inference{
    T \neq \top & \relate{\Gamma}{R}{\Gamma_r}
  }{
    \relate{(\Gamma,x\of T)}{R}{(\Gamma_r,x\of \llbracket T\rrbracket_R\,\kw{at}\,R(x))}
  }
  \\[2ex]
  \inference{
    \relate{\Gamma}{R}{\Gamma_r}
  }{
    \relate{(\Gamma,x\of \top)}{R}{\Gamma_r}
  }
  \quad
  \inference{
    \relate{\Gamma}{R}{\Gamma_r}
  }{
    \relate{(\Gamma,x\of\mathsf{copy}\,T)}{R}{(\Gamma_r, x \of \llbracket T\rrbracket_R)}
  }
\end{gather*}

\boxed{\toreg{\varphi}{R}}
\vspace{-15pt}
\[
  \toreg{\varphi}{R} = \{ R(x) \mid x \in \varphi \} 
\]

\boxed{\Gamma \vdash R}
\vspace{-15pt}
\[
  \forall x.\, (\exists T.\, \Gamma(x) = T) \text{ iff } (\exists \rho. \, R(x) = \rho)
\]
  \caption{Translation of \funk{} to a region calculus.}
  \label{fig:translate-to-regions}
\end{figure}

The translation from $\funk$ to this region calculus is type
preserving, that is, it maps well-typed programs to well-typed
programs. The proof is in the accompanying technical report.

\begin{theorem}[Translation to Regions Preserves Types]\ 
  \label{thm:trans-preserves-types}
  Suppose $\Gamma \vdash R$, $\Gamma \vdash \varphi_1$, and $\relate{\Gamma}{R}{\Gamma_r}$.
  \begin{enumerate}
  \item If $\Gamma; \varphi_1 \vdash e : T$, then $\Gamma_r; \toreg{\varphi_1}{R} \vdash
    \toreg{e}{R} : \toreg{T}{R}$.
  \item If $\Gamma; \varphi_1; \varphi_2 \vdash s : T$, then $\Gamma_r; \toreg{\varphi_1}{R} \vdash
    \toreg{s}{R} : \toreg{T}{R}$.
  \end{enumerate}
\end{theorem}

\noindent We already proved type safety in
Section~\ref{sec:type-safety} with respect to the direct semantics of
\funk. With Theorem~\ref{thm:trans-preserves-types} in hand, we could
write an alternative proof that relies on the type safety of the
region calculus.

The translation to regions is also be semantics preserving. That is,
evaluating a program $s$ under the direct semantics yields the same
result as translating the program and then evaluating under the
dynamic semantics of the region calculus~\cite{Pierce:2004fk}, for
which we use the name $\mathit{eval}_{\mathcal{R}}$. We have
tested this property on numerous programs and have found no
violations. The proof of this should be straightforward using a
simulation argument.

\begin{proposition}[Semantics Preserving]
  If $\emptyset;\emptyset; \emptyset \vdash s : T$, then
  $\mathit{eval}(s) = \mathit{eval}_{\mathcal{R}}(\toreg{s}{\emptyset})$.
\end{proposition}

\section{Related Work}
\label{sec:related-work}

Here we place the design of \funk{} in relation to other points in the
design space. To the best of our knowledge, \funk{} provides a unique
balance of safety, efficiency, and simplicity through its combination
of a regionless type and effect system and programmer controlled
function-local storage.

\paragraph{Efficient but unsafe} 
The lambda expressions of C++~\cite{ISO:2011uq,Jarvi:2007fk} give
programmers control over whether to capture references to variables or
to make copies.  The design of \funk{} offers the addition of static
checking to catch upward funargs with dangling references.

\paragraph{Safe but less efficient}
The blocks in Objective C~\cite{Apple:2011fk} give programmers a
choice between copying the value of a variable into function-local
storage (the default) or storing the variable on the heap (called
\textsf{\_\_block} storage) with automatic memory management.
Unfortunately, neither of these options is ideal for the most common
use case for first-class functions: downward funargs. In such
situations, storing the variable on the stack and capturing a
reference to it is both safe and the most efficient.

\paragraph{Safe but efficiency is less predictable}

Compilers for garbage-collected languages, such as Java, Standard ML,
and Scheme, optimize memory allocation by performing static analyses
(such as escape analysis~\cite{Goldberg:1990pi,Choi:1999bk}, region
inference~\cite{Tofte:1994uq}, and storage use
analysis~\cite{Serrano:1996uq}) to decide when objects can use a FIFO
memory management scheme.  However, from the programmer standpoint,
whether the static analyses succeeds on a given program is
unpredictable, which in turn means that the time and space efficiency
of the program is unpredictable~\cite{Tofte:2004fk}.  The design of
\funk{} offers an alternative in which programmers can mandate the
stack allocation of objects, thereby achieving predictable efficiency.

\paragraph{Safe and efficient, but complex}

Several programming languages employ type and effect systems to
provide safety and efficiency, but at the cost of exposing regions to
the programmer~\cite{Grossman:2002cr,Boyapati:2003fk}. (Several
intermediate representations also use explicit regions, but because
they are intermediate representations, regions are not necissarily
exposed to the programmer~\cite{Crary:1999fk,Tofte:1994uq}.)  Instead
of regions, the \funk{} design relies on traditional stack allocation
where parameters and local variables are implicitly allocated on the
stack. The effects of \funk{} are sets of variables instead of sets of
region names.

\section{Conclusion}
\label{sec:conclusion}

This paper presents a design for mixing first-class functions and
stack allocation that ensures type safety, enables the full range of
higher-order programming, and gives the programmer control over
efficiency while minimizing programmer-visible complexity.
The design uses a type and effect system based on variables instead of
region names. The system allows downward funargs and disallows upward
funargs with dangling references. To facilitate safe upward funargs,
the design gives programmers the choice of copying values into
function-local storage.
The paper formalizes the design in the \funk{} calculus, defining its
syntax, type system, and dynamic semantics, and gives a proof of type
safety. This paper also demonstrates that \funk{} is parametric with
respect to effects so effect abstractions can be implemented via
erasure. Finally, the paper relates \funk{} to a region calculus
through a type-directed translation.

\bibliographystyle{abbrvnat}
\bibliography{all}

\begin{thebibliography}{29}
\providecommand{\natexlab}[1]{#1}
\providecommand{\url}[1]{\texttt{#1}}
\expandafter\ifx\csname urlstyle\endcsname\relax
  \providecommand{\doi}[1]{doi: #1}\else
  \providecommand{\doi}{doi: \begingroup \urlstyle{rm}\Url}\fi

\bibitem[Aiken et~al.(1995)Aiken, F\"{a}hndrich, and Levien]{Aiken:1995fk}
A.~Aiken, M.~F\"{a}hndrich, and R.~Levien.
\newblock Better static memory management: improving region-based analysis of
  higher-order languages.
\newblock In \emph{Proceedings of the ACM SIGPLAN 1995 conference on
  Programming language design and implementation}, PLDI '95, pages 174--185,
  New York, NY, USA, 1995. ACM.
\newblock \doi{http://doi.acm.org/10.1145/207110.207137}.

\bibitem[Appel(1992)]{Appel:1992fk}
A.~W. Appel.
\newblock \emph{Compiling with continuations}.
\newblock Cambridge University Press, New York, NY, USA, 1992.
\newblock ISBN 0-521-41695-7.

\bibitem[{Apple Inc.}(2011)]{Apple:2011fk}
{Apple Inc.}
\newblock Blocks programming topics.
\newblock Technical report, Apple Inc., Cupertino, CA, March 2011.

\bibitem[Boyapati et~al.(2003)Boyapati, Salcianu, Beebee, and
  Rinard]{Boyapati:2003fk}
C.~Boyapati, A.~Salcianu, W.~Beebee, Jr., and M.~Rinard.
\newblock Ownership types for safe region-based memory management in real-time
  java.
\newblock In \emph{Proceedings of the ACM SIGPLAN 2003 conference on
  Programming language design and implementation}, PLDI '03, pages 324--337,
  New York, NY, USA, 2003. ACM.
\newblock \doi{http://doi.acm.org/10.1145/781131.781168}.

\bibitem[Cardelli(1997)]{Cardelli:1997fk}
L.~Cardelli.
\newblock \emph{Handbook of Computer Science and Engineering}, chapter Type
  Systems.
\newblock CRC Press, 1997.

\bibitem[Chamberlain et~al.(2011)Chamberlain, Deitz, Hoffswell, Plevyak, Zima,
  and Diaconescu]{Chamberlain:2005fd}
B.~Chamberlain, S.~Deitz, S.~Hoffswell, J.~Plevyak, H.~Zima, and R.~Diaconescu.
\newblock \emph{Chapel Specification}.
\newblock Cray Inc, 0.82 edition, October 2011.

\bibitem[Choi et~al.(1999)Choi, Gupta, Serrano, Sreedhar, and
  Midkiff]{Choi:1999bk}
J.-D. Choi, M.~Gupta, M.~Serrano, V.~C. Sreedhar, and S.~Midkiff.
\newblock Escape analysis for java.
\newblock In \emph{OOPSLA '99: Proceedings of the 14th ACM SIGPLAN conference
  on Object-oriented programming, systems, languages, and applications}, pages
  1--19, New York, NY, USA, 1999. ACM Press.

\bibitem[Crary et~al.(1999)Crary, Walker, and Morrisett]{Crary:1999fk}
K.~Crary, D.~Walker, and G.~Morrisett.
\newblock Typed memory management in a calculus of capabilities.
\newblock In \emph{Proceedings of the 26th ACM SIGPLAN-SIGACT symposium on
  Principles of programming languages}, POPL '99, pages 262--275, New York, NY,
  USA, 1999. ACM.
\newblock \doi{http://doi.acm.org/10.1145/292540.292564}.

\bibitem[de~Bruijn(1972)]{Bruijn:1972kx}
N.~de~Bruijn.
\newblock {Lambda calculus notation with nameless dummies, a tool for automatic
  formula manipulation, with application to the Church-Rosser theorem}.
\newblock \emph{Indagationes Mathematicae (Proceedings)}, 75\penalty0
  (5):\penalty0 381--392, 1972.
\newblock \doi{10.1016/1385-7258(72)90034-0}.

\bibitem[Flanagan et~al.(1993)Flanagan, Sabry, Duba, and
  Felleisen]{Flanagan:1993cg}
C.~Flanagan, A.~Sabry, B.~F. Duba, and M.~Felleisen.
\newblock The essence of compiling with continuations.
\newblock In \emph{PLDI '93: Proceedings of the ACM SIGPLAN 1993 conference on
  Programming language design and implementation}, pages 237--247, New York,
  NY, USA, 1993. ACM Press.

\bibitem[Fluet and Morrisett(2006)]{Fluet:2006hb}
M.~Fluet and G.~Morrisett.
\newblock Monadic regions.
\newblock \emph{J. Funct. Program.}, 16\penalty0 (4-5):\penalty0 485--545,
  2006.

\bibitem[Goldberg and Park(1990)]{Goldberg:1990pi}
B.~Goldberg and Y.~G. Park.
\newblock Higher order escape analysis.
\newblock In \emph{ESOP}, 1990.

\bibitem[Grossman et~al.(2002)Grossman, Morrisett, Jim, Hicks, Wang, and
  Cheney]{Grossman:2002cr}
D.~Grossman, G.~Morrisett, T.~Jim, M.~Hicks, Y.~Wang, and J.~Cheney.
\newblock Region-based memory management in cyclone.
\newblock In \emph{PLDI '02: Proceedings of the ACM SIGPLAN 2002 Conference on
  Programming language design and implementation}, pages 282--293, New York,
  NY, USA, 2002. ACM Press.

\bibitem[Helsen and Thiemann(2000)]{Helsen:2000fk}
S.~Helsen and P.~Thiemann.
\newblock Syntactic type soundness for the region calculus.
\newblock In \emph{4th International Workshop on Higher Order Operational
  Techniques in Semantics (HOOTS 2000)}, volume~41 of \emph{ENTCS}, pages
  1--19. Elsevier, 2000.

\bibitem[Henglein et~al.(2001)Henglein, Makholm, and Niss]{Henglein:2001zr}
F.~Henglein, H.~Makholm, and H.~Niss.
\newblock A direct approach to control-flow sensitive region-based memory
  management.
\newblock In \emph{Proceedings of the 3rd ACM SIGPLAN international conference
  on Principles and practice of declarative programming}, PPDP '01, pages
  175--186, New York, NY, USA, 2001. ACM.
\newblock \doi{http://doi.acm.org/10.1145/773184.773203}.

\bibitem[ISO(2011)]{ISO:2011uq}
ISO.
\newblock Working draft, standard for programming language {C++}.
\newblock Technical Report N3242, ISO, February 2011.

\bibitem[J{\"a}rvi et~al.(2007)J{\"a}rvi, Freeman, and Crowl]{Jarvi:2007fk}
J.~J{\"a}rvi, J.~Freeman, and L.~Crowl.
\newblock Lambda expressions and closures for c++ (revision 1).
\newblock Technical Report N2329, ISO/IEC JTC 1 SC22 WG21, June 2007.

\bibitem[Miller and Rozas(1994)]{Miller:1994fk}
J.~S. Miller and G.~J. Rozas.
\newblock Garbage collection is fast, but a stack is faster.
\newblock AI Memos AIM-1462, MIT Artificial Intelligence Lab, March 1994.

\bibitem[Moses(1970)]{Moses:1970fk}
J.~Moses.
\newblock The function of {FUNCTION} in {LISP} or why the {FUNARG} problem
  should be called the environment problem.
\newblock \emph{SIGSAM Bull.}, pages 13--27, July 1970.
\newblock \doi{http://doi.acm.org/10.1145/1093410.1093411}.

\bibitem[Pierce(2002)]{Pierce:2002hj}
B.~C. Pierce.
\newblock \emph{Types and {P}rogramming {L}anguages}.
\newblock MIT Press, 2002.

\bibitem[Pierce(2004)]{Pierce:2004fk}
B.~C. Pierce, editor.
\newblock \emph{Advanced Topics in Types and Programming Languages}.
\newblock The MIT press, 2004.

\bibitem[Plotkin(1975)]{G.-D.-Plotkin:1975on}
G.~D. Plotkin.
\newblock Call-by-name, call-by-value and the lambda-calculus.
\newblock \emph{Theoretical Computer Science}, 1\penalty0 (2):\penalty0
  125--159, December 1975.

\bibitem[Sabry(2002)]{sabry02:_minml}
A.~Sabry.
\newblock {MinML}: Syntax, static semantics, dynamic semantics, and type
  safety.
\newblock Course notes for b522, February 2002.

\bibitem[Serrano and Feeley(1996)]{Serrano:1996uq}
M.~Serrano and M.~Feeley.
\newblock Storage use analysis and its applications.
\newblock In \emph{Proceedings of the first ACM SIGPLAN international
  conference on Functional programming}, ICFP '96, pages 50--61, New York, NY,
  USA, 1996. ACM.
\newblock \doi{http://doi.acm.org/10.1145/232627.232635}.

\bibitem[Steele(1978)]{Guy-L.-Steele:1978yq}
G.~L. Steele.
\newblock Rabbit: A compiler for {Scheme}.
\newblock Technical report, Cambridge, MA, USA, 1978.

\bibitem[Talpin and Jouvelot(1992)]{Talpin:1992vn}
J.-P. Talpin and P.~Jouvelot.
\newblock The type and effect discipline.
\newblock In \emph{Logic in Computer Science, 1992. LICS '92., Proceedings of
  the Seventh Annual IEEE Symposium on}, pages 162 --173, jun 1992.
\newblock \doi{10.1109/LICS.1992.185530}.

\bibitem[Tofte and Talpin(1994)]{Tofte:1994uq}
M.~Tofte and J.-P. Talpin.
\newblock Implementation of the typed call-by-value lambda-calculus using a
  stack of regions.
\newblock In \emph{POPL '94: Proceedings of the 21st ACM SIGPLAN-SIGACT
  symposium on Principles of programming languages}, pages 188--201. ACM Press,
  1994.

\bibitem[Tofte and Talpin(1997)]{Tofte:1997fk}
M.~Tofte and J.-P. Talpin.
\newblock Region-based memory management.
\newblock \emph{Inf. Comput.}, 132\penalty0 (2):\penalty0 109--176, 1997.

\bibitem[Tofte et~al.(2004)Tofte, Birkedal, Elsman, and
  Hallenberg]{Tofte:2004fk}
M.~Tofte, L.~Birkedal, M.~Elsman, and N.~Hallenberg.
\newblock A retrospective on region-based memory management.
\newblock \emph{Higher-Order and Symbolic Computation Journal}, 17:\penalty0
  245--265, 2004.

\end{thebibliography}

\end{document}